\documentclass[%
 print,
 amsmath,amssymb,
 aps,
 pra,
]{revtex4-2}

\usepackage{graphicx}% Include figure files
\usepackage{dcolumn}% Align table columns on decimal point
\usepackage{bm}% bold math
\usepackage{physics}
\usepackage{hyperref} % 交叉引用跳转
\hypersetup{
    colorlinks=true,      % 链接显示为彩色文字（非边框）
    linkcolor=blue,       % 内部链接颜色
    citecolor=red,        % 文献引用颜色
    urlcolor=cyan,        % URL 颜色
    pdfborder={0 0 0},    % 取消链接边框
}
\usepackage[mathlines]{lineno}% Enable numbering of text and display math
\usepackage{enumerate}

\begin{document}

\preprint{APS/123-QED}

\title{Inverse-squeezing receivers for squeezed-state pulse-position modulation under ideal and phase-diffusion conditions}

\author{Enhao Bai}
  \affiliation{Information Support Force Engineering University, Wuhan 430035, China}
  \affiliation{School of Computer, Electronics and Information, Guangxi University, Nanning 530006, China}
\author{Fengkai Sun}
  \affiliation{School of Computer, Electronics and Information, Guangxi University, Nanning 530006, China}

\author{Tianyi Wu}
  % \email{18142630162@163.com}
  \affiliation{Information Support Force Engineering University, Wuhan 430035, China}
\author{Huankai Zhang}
  \affiliation{Information Support Force Engineering University, Wuhan 430035, China}
\author{Jian Peng}%
  \email{pengjian@nudt.edu.cn}
  \affiliation{Information Support Force Engineering University, Wuhan 430035, China}
\author{Chen Dong}
  \email{dongchengfkd@163.com}
  \affiliation{Information Support Force Engineering University, Wuhan 430035, China}

\author{Zhenrong Zhang}
  \email{zzr76@gxu.edu.cn}
  \affiliation{School of Computer, Electronics and Information, Guangxi University, Nanning 530006, China}

\date{\today}

\begin{abstract}
We introduce a squeezed-state pulse-position modulation (S-PPM) format, where the empty slots are squeezed vacuum states and the pulse slot is a displaced squeezed state.
Based on this property, we propose an inverse-squeezing conditional pulse-nulling (IS-CPN) receiver. In the ideal case, inverse squeezing maps S-PPM into an equivalent coherent-state PPM signal with a large pulse energy, leading to a closed-form expression for the receiver error probability.
We further analyze IS-CPN under common phase diffusion using a finite-path MAP formulation with phase-averaged likelihoods. Numerical results show that IS-CPN outperforms conventional CPN under the same energy constraint and remains advantageous under phase noise and finite photon-number resolution. These results demonstrate that combining squeezed-state modulation with inverse-squeezing conditional nulling can improve photon-efficient optical communication.
\end{abstract}

\maketitle

%\tableofcontents
\section{Introduction}
Pulse-position modulation (PPM) is a fundamental modulation format for photon-efficient optical communication. In an M-ary PPM symbol, information is encoded in the temporal position of a single optical pulse among M possible slots. Since only one slot carries the pulse while the remaining slots are empty, PPM is especially attractive in the low-photon-number regime. For coherent-state PPM (C-PPM), ideal direct detection provides a simple and widely used benchmark--Standard Quantum Limit (SQL)\cite{review_2021}, while the ultimate minimum error probability is given by the Helstrom bound \cite{Helstrom}. Between these two limits, structured quantum receivers such as conditional pulse-nulling (CPN) receivers \cite{Dolinar_PPM} improve performance by combining displacement operations, photon counting, and sequential decision rules.

In recent years, the encoded transmission of squeezed states has seen some development, such as squeezed-state on-off keying (S-OOK) \cite{paris_2001}, squeezed-state binary phase shift keying (S-BPSK) \cite{Olivares_2018}, and PPM using a combination of vacuum and squeezed states (where the squeezed state represents the pulse and the vacuum state represents the empty slot) \cite{Wang_PPM}. Since the overlap between squeezed states can be reduced by adjusting the squeezing strength and phase, the Helstrom limit for squeezed-state encoding is generally lower than that for coherent-state encoding with the same energy.
In particular, Bai. et al. recently proposed an IS-Kennedy receiver for discriminate S-BPSK signals, which innovatively employs inverse squeezing (IS) to convert S-BPSK into a higher-energy C-BPSK signal, thereby achieving a lower error probability \cite{IS-Kennedy}.

Motivated by these ideas, we introduce in this work a squeezed-state pulse-position modulation (S-PPM) format. In the proposed S-PPM alphabet, the empty slots are encoded as squeezed vacuum states, while the pulse slot is encoded as a displaced squeezed state.
The use of squeezed states in a PPM alphabet also changes the choice of appropriate benchmarks and receivers. In ordinary C-PPM, the empty slots are vacuum states and are therefore photon-counting silent; direct detection is consequently a natural standard quantum limit. In the proposed S-PPM format, however, the empty slots are squeezed vacuum states, which can produce photon counts even in the absence of displacement. Direct detection therefore suffers from intrinsic false alarms and is no longer a unique SQL benchmark for S-PPM. 
Homodyne detection is also a natural classical benchmark because the two single-slot states have the same covariance matrix and differ only in their mean displacement. However, we have also observed that, for weak signals, the error probability of direct detection is actually lower than that of homodyne detection. Therefore, throughout this work, we report DD and homodyne detection as separate classical receiver benchmarks rather than defining a single S-PPM SQL.

To exploit the structure of the proposed S-PPM alphabet, we further propose an inverse-squeezing conditional pulse-nulling (IS-CPN) receiver. The receiver first applies a slotwise inverse-squeezing operation and then performs a sequential conditional pulse-nulling measurement with photon-number-resolving detection. In the ideal phase-matched case, the inverse-squeezing operation exactly converts the S-PPM codeword into an equivalent coherent-state PPM codeword with large pulse energy. This allows the receiver to inherit the structure of conventional CPN while operating on an enhanced effective energy.

A key practical issue is phase sensitivity. Since squeezed states depend on a phase reference, channel phase diffusion or phase-reference mismatch can rotate the squeezing ellipse and reduce the benefit of inverse squeezing. We therefore analyze the proposed S-PPM and IS-CPN receiver not only under ideal phase matching, but also under a common phase-diffusion model in which all slots within one PPM symbol share the same random phase realization. Under phase diffusion, inverse squeezing no longer produces an ordinary coherent-state PPM signal; a residual Gaussian unitary remains after the inverse-squeezing operation. To evaluate the receiver in this regime, we develop a finite-path maximum-a-posteriori formulation based on thresholded photon-number records and phase-averaged likelihoods.

The rest of this paper is organized as follows. Section II introduces the proposed S-PPM signal model and derives the ideal benchmarks. Section III presents the IS-CPN receiver and its ideal error probability. Section IV analyzes the phase-diffusion model and the finite-path MAP evaluation of IS-CPN. Section V gives numerical results and discussion. Section VI concludes the paper.

\section{Signal model, quantum bound, and classical receivers for M-ary Squeezed-state PPM}
In this section, we consider a binary single-slot alphabet in which the symbols 0 and 1 are encoded as a squeezed vacuum state (SVS) and a displaced squeezed state (DSS), respectively,
\begin{equation}
  \text{symbol `0': }\ket{0,r} = S(r)\ket{0},\quad
  \text{symbol `1': }\ket{\alpha,r} = D(\alpha)S(r)\ket{0},
\end{equation}
where $S(r)$ and $D(\alpha)$ denote the squeezing and displacement operators, respectively. Throughout this work, we assume $\alpha\in \mathbb{R}_+$ and $r\in \mathbb{R}_+$. Under this convention, the two symbols have the same covariance matrix, and the discrimination information is carried entirely by their mean displacement.

The $M$-ary S-PPM codebook is constructed in the usual PPM manner: each codeword contains exactly one displaced squeezed state and $M-1$ squeezed-vacuum slots. Denoting the $m$-th hypothesis by $H_m$ with $m=1,\cdots, M$, the corresponding codeword is 
\begin{equation}
  \ket{\Psi_m} = \bigotimes_{k=1}^{M} \ket{\psi_{m,k}},\quad 
  \ket{\psi_{m,k}} = \left\{\begin{aligned}
    &\ket{\alpha,r},\ k=m,\\
    &\ket{0,r},\ k\neq m.
  \end{aligned}\right.
  \label{eq:s_ppm}
\end{equation}
The mean photon numbers of the two slot symbols are
\begin{equation}
  N_0=\sinh^2 r,\quad N_1=\alpha^2+\sinh^2 r,
\end{equation}
so that the total mean photon number per M-ary S-PPM symbol is
\begin{equation}
  N = (M-1)\cdot N_0+N_1 = \alpha^2 + M\sinh^2 r.
\end{equation}
To compare S-PPM fairly with C-PPM under the same total-energy constraint, we introduce the squeezing-energy fraction
\begin{equation}
  \beta \triangleq \frac{N_{\rm sq}}{N}=\frac{M\sinh^2 r}{\alpha^2+M\sinh^2 r},
\end{equation}
where $N_\text{sq}$ denotes the total energy associated with the $M$ squeezing operations.

\subsection{Ideal Helstrom bound}

We first derive the minimum error probability for discriminating the $M$-ary S-PPM codewords in the ideal pure-state case. The codewords form an equiprobable geometrically uniform pure-state set. Therefore, the square-root measurement (SRM) attains the Helstrom bound, and the problem is determined by the pairwise overlap between two distinct codewords.

For $i\neq j$, the two PPM codewords differ in exactly two slots. Hence their overlap is
\begin{equation}
  \label{eq:gamma}
  \begin{aligned}
    \Gamma &= \braket{\Psi_i}{\Psi_j}
    =\left|\braket{0,r}{\alpha,r}\right|^2
    = \left|\matrixel{0}{S^\dagger(r)D(\alpha)S(r)}{0}\right|^2\\
    &=\left|\matrixel{0}{D(\alpha \text{e}^r)}{0}\right|^2 = \left|\matrixel{0}{D(\gamma)}{0}\right|^2
    = \exp\left(-N_\text{eff}\right),
  \end{aligned}
\end{equation}
where $\gamma \equiv \alpha e^r$ and $N_\text{eff} \equiv \left|\gamma\right|^2$. 
The Gram matrix of the $M$-state ensemble has the form
\begin{equation}
  \text{Gram} = (1-\Gamma)\cdot\mathbb{I}_M + \Gamma\cdot \mathbb{J}_M
\end{equation}
where $\mathbb{I}_M$ is the $M\times M$ identity matrix and $\mathbb{J}_M$ is the $M\times M$ all-ones matrix. Its eigenvalues are
\begin{equation}
  \lambda_1 = 1 + (M-1)\Gamma,\quad \lambda_2 = 1 - \Gamma\ (\text{with multiplicity} M-1)
\end{equation}
Substituting these eigenvalues into the standard SRM expression gives the ideal Helstrom bound
\begin{equation}
  \label{eq:hb}
  P_{\rm HB}^{\rm S}
  =1-\frac{1}{M^2}\left[\sqrt{1+(M-1)\Gamma}+(M-1)\sqrt{1-\Gamma}\right]^2.
\end{equation}
Since $\Gamma = \exp(-N_\text{eff})$, the ideal Helstrom bound depends on $\alpha$ and $r$ only through $N_\text{eff}$. Thus, under a fixed total-energy constraint, minimizing the ideal Helstrom bound is equivalent to maximizing $N_\text{eff}$. Then the resulting optimal squeezing-energy fraction is
\begin{equation}
  \label{eq:ideal_opt_beta}
  \beta_\text{opt}^\text{ideal} (N) = \frac{N}{M + 2N}.
\end{equation}
At this operating point, 
\begin{equation}
  \alpha^2 = \frac{N(N+M)}{2N+M},\quad \text{e}^{2r} = \frac{2N+M}{M},\quad N_\text{eff} = N\left(1 + \frac{N}{M}\right)
\end{equation}

\subsection{Direct-detection and homodyne benchmarks for S-PPM}
We next introduce two classical receiver benchmarks for the proposed S-PPM alphabet. First, we define the single-slot binary test
\begin{equation}
  H_0':\ \ket{0,r},\quad H_1':\ \ket{\alpha,r}.
\end{equation}
Then, for the ideal on--off direct detection (DD), 
\begin{equation}
  \begin{aligned}
    &u_0 = p\left(n=0|H_0'\right) = \left|\braket{0}{0,r}\right|^2 = \frac{1}{\cosh r},\\
    &v_0 = p\left(n\ne 0|H_0'\right) = 1 - u_0,\\
    &u_1 = p\left(n=0|H_1'\right) = \left|\braket{0}{\alpha,r}\right|^2 = \frac{1}{\cosh r}\exp\left[-\alpha^2\left(1 + \tanh r\right)\right],\\
    &v_1 = p\left(n\ne 0|H_1'\right) = 1 - u_1.
  \end{aligned}
\end{equation}
Since $u_1 < u_0$ for any $\alpha>0$, one has $v_1 > v_0$. Therefore, under equal priors, the MAP rule reduces to the following intuitive decision strategy: if at least one slot clicks, choose uniformly among the clicked slots; if no slot clicks, choose uniformly among all $M$ hypotheses.

Without loss of generality, suppose the true hypothesis is $H_1$, so that the first slot contains $|\alpha,r\rangle$ and the remaining $M-1$ slots contain $|0,r\rangle$. If the pulse slot clicks and exactly $k$ of the $M-1$ empty slots also click, then the corresponding event probability is
\begin{equation}
  v_1 \binom{M-1}{k} v_0^k u_0^{M-1-k},
\end{equation}
and the correct-decision probability conditioned on this event is $1/(k+1)$, since there are $k+1$ clicked slots in total. If the pulse slot does not click and at least one empty slot clicks, an error is always made. The only no-click event that can still lead to a correct decision is the all-zero pattern, which occurs with probability $u_1u_0^{M-1}$ and yields a correct guess with probability $1/M$. Hence, the overall correct-decision probability is
\begin{equation}
  \begin{aligned}
    P_{\text{cor}}^\text{DD,S}
    &= v_1\sum_{k=0}^{M-1}\binom{M-1}{k}\frac{v_0^k u_0^{M-1-k}}{k+1}
    +\frac{1}{M}u_1u_0^{M-1}\\
    &= \frac{v_1}{M v_0}\left(1-u_0^M\right) + \frac{1}{M}u_1u_0^{M-1}\\
    &= \frac{1-u_1}{M(1-u_0)}\left(1-u_0^M\right) + \frac{1}{M}u_1u_0^{M-1}.
  \end{aligned}
\end{equation}
Therefore, the symbol error probability of ideal direct detection for $M$-ary S-PPM is
\begin{equation}
  P_{\text{err}}^\text{DD,S}=1-P_{\text{cor}}^\text{DD,S}.
\end{equation}

For the homodyne, its outcomes obey two Gaussian distributions with the same variance,
\begin{equation}
  p\left(x|H_0'\right)=\frac{1}{\sqrt{2\pi V_x}}\exp\!\left(-\frac{x^2}{2V_x}\right),\quad
  p\left(x|H_1'\right)=\frac{1}{\sqrt{2\pi V_x}}\exp\!\left(-\frac{(x-\alpha)^2}{2V_x}\right),
\end{equation}
where $V_x=\frac{1}{4}e^{-2r}$.
For equal priors, the optimal threshold is $x=\alpha/2$. The false-alarm and miss probabilities are then identical and are given by
\begin{equation}
  \epsilon
  =Q\!\left(\frac{\alpha/2}{\sqrt{V_x}}\right)
  =Q(\alpha e^{r})=Q\left(\sqrt{N_\text{eff}}\right)
  =\frac{1}{2}\,\mathrm{erfc}\!\left(\sqrt{\frac{N_\text{eff}}{2}}\right)
  \label{eq:slot_error_homodyne}
\end{equation}
where $Q(\cdot)$ is the standard Gaussian Q-function.

Then we adopt the following classical post-processing rule: if exactly one slot is declared as the pulse slot, that slot is selected. If more than one slot is declared as the pulse slot, one of the declared slots is chosen uniformly at random. If no slot is declared as the pulse slot, one of the 
$M$ hypotheses is chosen uniformly at random.
Assume without loss of generality that the true pulse is in a given slot.
The true slot is correctly declared as a pulse with probability $1 - \epsilon$.
If exactly $k$ of the remaining $M-1$ empty slots are falsely declared as pulses, the correct hypothesis is selected with probability $1/(k+1)$.
In addition, if the true pulse slot is missed and all empty slots are also declared empty, the all-zero outcome leads to a random guess among all $M$ hypotheses. Therefore, the correct-decision probability is
\begin{equation}
  \begin{aligned}
    P_{\rm cor}^{\rm Hom,S}&=(1-\epsilon) \sum_{k=0}^{M-1} \binom{M-1}{k} \epsilon^{k}(1-\epsilon)^{M-1-k} \frac{1}{k+1}+\frac{1}{M} \epsilon(1-\epsilon)^{M-1}\\
    &= \frac{1-\epsilon}{M\epsilon} \left[ 1-(1-\epsilon)^{M} \right] + \frac{\epsilon}{M} (1-\epsilon)^{M-1}.
  \end{aligned}
\end{equation}
The homodyne benchmark of M-ary S-PPM is therefore
\begin{equation}
  P_{\rm err}^{\rm Hom,S}
  =1 - P_{\rm cor}^{\rm Hom,S}.
  \label{eq:sql_s-ppm}
\end{equation}

\begin{figure}[htbp]
  \centering
  \includegraphics[scale=1.0]{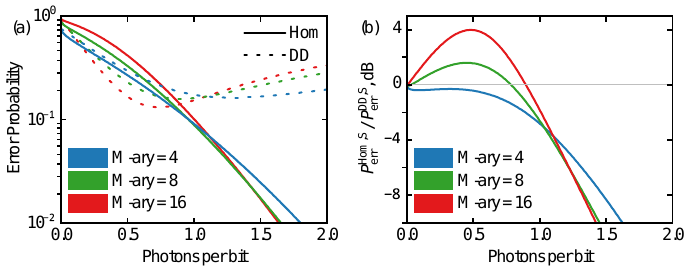}
  \caption{The DD and Hom error probabilities of S-PPM signals of different modulations. Negative values indicate that homodyne outperforms DD, while positive values indicate that DD outperforms homodyne.}
  \label{fig:ratio_hom_dd}
\end{figure}

Figure~\ref{fig:ratio_hom_dd}(a) compares the DD and homodyne error probabilities of S-PPM for different PPM orders. As the signal strength increases, the DD error probability may stop decreasing and eventually rise. This behavior is caused by the squeezed-vacuum empty slots. Under the optimized energy allocation, increasing the total energy can increase the squeezing parameter ($r$), which decreases the no-click probability ($u_0=1/\cosh r$) and therefore increases the empty-slot click probability ($v_0=1-u_0$). The resulting false alarms degrade the DD performance. Figure~\ref{fig:ratio_hom_dd}(b) shows the ratio ($P_\text{err}^\text{Hom,S} / P_\text{err}^\text{DD,S}$) in dB, where negative values indicate that homodyne detection outperforms DD, while positive values indicate that DD outperforms homodyne detection. In most of the plotted parameter region, homodyne detection gives a lower error probability than DD. However, DD can outperform homodyne detection in certain weak-signal regimes. This behavior confirms that assigning a unique SQL to the S-PPM alphabet is ambiguous. Therefore, we treat DD and homodyne detection as separate classical receiver benchmarks rather than defining a single S-PPM SQL.

\section{Inverse-squeezing conditional pulse nulling receiver for S-PPM}
In this section, we introduce an inverse-squeezing conditional pulse-nulling receiver for the M-ary S-PPM alphabet defined in Sec.~II.
Throughout this section, we consider the ideal phase-matched case, in which the receiver has perfect knowledge of the squeezing parameter and phase reference, and the photon-number-resolving detector is assumed to be ideal, with unit efficiency, no dark counts, and no background photons.

\begin{figure}[htbp]
  \centering
  \includegraphics[scale=1.0]{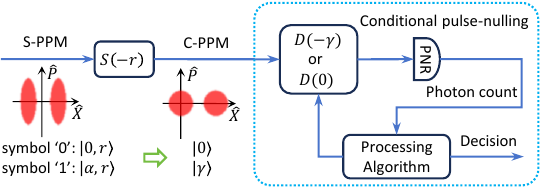}
  \caption{The structure of inverse-squeezing conditional pulse nulling receiver, where $\gamma = \alpha e^r$.}
  \label{fig:receiver_structure}
\end{figure}

The proposed receiver consists of two stages. The first stage applies a slotwise inverse-squeezing operation, which converts the S-PPM codeword into an equivalent coherent-state PPM codeword. 
The second stage performs a sequential conditional pulse-nulling measurement followed by a maximum-likelihood decision rule. The receiver structure is illustrated in Fig.~\ref{fig:receiver_structure}.

\subsection{Inverse-squeezing transformation}
The two elementary states of the S-PPM alphabet are the squeezed vacuum state $\ket{0, r} = S(r)\ket{0}$ and the displaced squeezed state $\ket{\alpha, r} = D(\alpha)S(r)\ket{0}$. In the ideal receiver, an inverse-squeezing operation $S(-r)$ is applied to each temporal slot. 
Using $S(-r)D(\alpha)S(r) = D(\alpha e^r)$, for real $\alpha$ and real $r$, the two slot states are transformed as
\begin{equation}
  \ket{0,r} \overset{S(-r)}{\longrightarrow} \ket{0},\ \ket{\alpha,r} \overset{S(-r)}{\longrightarrow} \ket{\gamma},
  \label{eq:IS_transform}
\end{equation}
where $\gamma = \alpha e^r$.
Therefore, after the inverse-squeezing front end, the $m$-th S-PPM codeword (Eq.~\eqref{eq:s_ppm}) is mapped to the coherent-state PPM codeword
\begin{equation}
  \ket{\Phi_m} = \bigotimes_{k=1}^{M} \ket{\phi_{m,k}},\quad 
  \ket{\phi_{m,k}} = \left\{\begin{aligned}
    &\ket{\gamma},\ k=m,\\
    &\ket{0},\ k\neq m.
  \end{aligned}\right.
\end{equation}
Thus, in the ideal phase-matched case, the inverse-squeezing module does not merely improve the subsequent detection stage; it exactly converts the S-PPM discrimination problem into a C-PPM discrimination problem with pulse energy
\begin{equation}
  N_\text{eff} = \left|\gamma\right|^2 = \alpha^2 e^{2r}.
\end{equation}
This effective energy is the same parameter that appears in the Helstrom bound and the homodyne benchmark derived in Sec.~II.

\subsection{Conditional pulse-nulling measurement}
After the inverse-squeezing transformation, the receiver processes the equivalent C-PPM codeword using a conditional pulse-nulling strategy. The receiver starts from the first temporal slot and operates sequentially.

In the nulling mode, denoted by $N$, the receiver applies the displacement $D(-\gamma)$ to the currently tested slot. If the current slot actually contains the pulse state $\ket{\gamma}$, the displacement maps it to the vacuum state. If the current slot is empty, the displacement maps the vacuum state to the coherent state $\ket{-\gamma}$. The displaced state is then measured by a photon-number-resolving detector.

Let $n_k$ be the detected photon number in the $k$-th slot. For a given threshold $n_{th} \ge 1$, we define the binary detection outcome 
\begin{equation}
  y_k = \left\{\begin{aligned}
    &0,\ n_k< n_{th},\\
    &1,\ n_k\ge n_{th},
  \end{aligned}\right.
\end{equation}
where $y_k = 0$ represents a thresholded no-click event and $y_k = 1$ represents a thresholded click event.

The receiver uses the following sequential rule. It remains in the nulling mode as long as click events are observed. Once the first no-click event occurs, the receiver switches to the direct-detection mode, denoted by $D$, for all remaining slots. In the direct-detection mode, no additional displacement is applied.

\begin{figure}[htbp]
  \centering
  \includegraphics[scale=1.0]{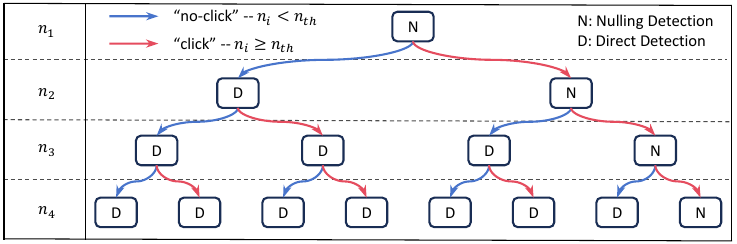}
  \caption{The decision tree of ideal inverse-squeezing conditional pulse-nulling receiver ($M=4$).}
  \label{fig:decision_tree}
\end{figure}

Equivalently, the operation applied to the $k$-th slot is
\begin{equation}
  u_k\left(y_{<k}\right) = \left\{\begin{aligned}
    &N,\ k=1,\\
    &N,\ y_1=\cdots=y_{k-1}=1,\\
    &D,\ \text{otherwise},
  \end{aligned}\right.
\end{equation}
where $y_{<k} = (y_1,\cdots,y_{k-1})$. This sequential measurement tree is shown in Fig.~\ref{fig:decision_tree}.

The final decision can be viewed as a maximum-likelihood decision based on the complete thresholded record $\mathbf{y} = (y_1,\cdots,y_M)$. In the ideal case, this rule has a simple operational interpretation: if a click is observed after the receiver has switched to direct detection, the corresponding slot is selected as the pulse position; otherwise, the first no-click slot is selected. This rule is equivalent to the maximum-likelihood rule under the ideal C-PPM model described above.

\subsection{Ideal error probability}
We now derive the error probability of the ideal IS-CPN receiver. For a coherent state $\ket{\gamma}$, the probability that the photon-number-resolving detector produces a thresholded no-click event is
\begin{equation}
  q_T = \Pr\left\{n<n_{th}|\ket{\gamma}\right\} = e^{-N_\text{eff}} \sum_{n=0}^{n_{th}-1} \frac{(N_\text{eff})^n}{n!}.
  \label{eq:no_click_prob_ideal}
\end{equation}
Equivalently, the slotwise thresholded probabilities $p\left\{y_k|x_k^{\left(m\right)},u_k\right\}$ can be written as
\begin{equation}
  \begin{array}{ll}
    p\left\{0|0,N\right\} = \Pr\left\{n<n_{th}|D(-\gamma)\ket{0}\right\} = q_T,& p\left\{1|0,N\right\} = 1 - q_T,\\
    p\left\{0|1,N\right\} = \Pr\left\{n<n_{th}|D(-\gamma)\ket{\gamma}\right\} = 1,& p\left\{1|1,N\right\} = 0\\
    p\left\{0|0,D\right\} = \Pr\left\{n<n_{th}|\ket{0}\right\} = 1,& p\left\{1|0,D\right\} = 0,\\
    p\left\{0|1,D\right\} = \Pr\left\{n<n_{th}|\ket{\gamma}\right\} = q_T,& p\left\{1|1,D\right\} = 1 - q_T,
  \end{array}
\end{equation}
where $x_k^{\left(m\right)} = \left\{\begin{aligned}
  &1,\ k=m,\\
  &0,\ k\neq m
\end{aligned}\right.$ is the slot-type indicator of hypothesis $H_m$.

Assume that the true pulse position is $m$. If all of the first $m-1$ empty slots generate clicks under nulling, the receiver continues nulling until the $m$-th slot. The $m$-th pulse slot is then exactly nulled to the vacuum and produces a no-click event with probability one, leading to the correct decision.
An error can occur only if the receiver switches to direct detection too early. Suppose that the first erroneous no-click occurs in the $t$-th slot, where $t<m$. This event has probability
\begin{equation}
  \left(p\left\{1|0,N\right\}\right)^{t-1} \cdot p\left\{0|0,N\right\} = \left(1 - q_T\right)^{t-1} q_T.
\end{equation}
After this premature switch, all empty slots between $t+1$ and $m-1$ remain vacuum under direct detection and therefore produce no clicks. The receiver can still recover the correct pulse position if the true pulse slot $m$ produces a click under direct detection. Hence an error occurs only when the true pulse slot also produces a no-click event, which has probability $p\left\{0|1,D\right\} = q_T$. Therefore, conditioned on hypothesis $H_m$, the error probability is
\begin{equation}
  \begin{aligned}
    P_{\text{err}|\text{m}} &= \left\{\sum_{t=1}^{m-1}\left(p\left\{1|0,N\right\}\right)^{t-1} \cdot p\left\{0|0,N\right\}\right\}\cdot p\left\{0|1,D\right\}\\
    &= \sum_{t=1}^{m-1}\left(1 - q_T\right)^{t-1} q_T^2 = q_T\left[1 - \left(1 - q_T\right)^{m-1}\right].
  \end{aligned}
\end{equation}
Averaging over the $M$ equiprobable pulse positions, we obtain
\begin{equation}
  P_\text{err}^\text{IS-CPN}(q_T) = \frac{1}{M} \sum_{m=1}^{M} q_T\left[1 - \left(1 - q_T\right)^{m-1}\right] = \frac{1}{M}\left[(1 - q_T)^M + M\cdot q_T - 1\right].
\end{equation}
Since
\begin{equation}
  \frac{\text{d}}{\text{d}q_T}P_\text{err}^\text{IS-CPN}(q_T) = 1 - (1 - q_T)^{M-1} \ge 0,
\end{equation}
the error probability is a monotonically increasing function of $q_T$. Therefore, the optimal threshold in the ideal case is
\begin{equation}
  n_{th} = 1,
\end{equation}
for which $q_T = e^{-N_\text{eff}}$.
Consequently, the ideal symbol error probability of the IS-CPN receiver is
\begin{equation}
  P_\text{err,ideal}^\text{IS-CPN} = \frac{1}{M} \left[\left(1 - \text{e}^{-N_\text{eff}}\right)^M + M\text{e}^{-N_\text{eff}} - 1\right]
  \label{eq:is-cpn_ideal}
\end{equation}
This result has the same functional form as the conventional CPN receiver for C-PPM, but with the coherent pulse energy replaced by the inverse-squeezing-enhanced energy $N_\text{eff}$.

We define the energy gain provided by the IS module as:
\begin{equation}
  G(N,\beta,M) = 10\cdot \log_{10} \left(\frac{N_\text{eff}}{N}\right)\ (\rm dB),
\end{equation}
which, under the optimal allocation, reduces to 
\begin{equation}
  G(N,M)\vert_{\beta = \beta_\text{opt}^\text{ideal}} = 10\cdot \log_{10} \left(1 + \frac{N}{M}\right)\ (\rm dB).
\end{equation}
This gain explains why the IS-CPN receiver can outperform the conventional CPN receiver under the same total-energy constraint. The inverse-squeezing operation converts part of the transmitted squeezing energy into an effective displacement enhancement, thereby allowing the conditional pulse-nulling stage to operate on a brighter equivalent coherent-state PPM signal.

\section{The performance of IS-CPN in the presence of phase diffusion}
\subsection{channel phase--diffusion model}
To investigate the robustness of the proposed scheme, we now consider a channel phase-diffusion model, in which the optical field acquires a random phase during propagation \cite{phase_diffusion_01, Olivares_phase_diffusion,IS-Kennedy,Olivares_2018}.
We assume that, within one symbol interval, the phase fluctuation is slow enough that all $M$ temporal slots experience the same random phase realization $\phi$, while $\phi$ varies independently from symbol to symbol. This common-phase model is physically appropriate when the symbol duration is much shorter than the laser coherence time or the characteristic timescale of the channel phase fluctuation.

The random phase is assumed to follow a zero-mean distribution $p_{\sigma}(\phi)$ characterized by the phase-diffusion strength $\sigma$ \cite{phase_diffusion_01}. For definiteness, one may adopt the Gaussian model \cite{Olivares_phase_diffusion,IS-Kennedy,Olivares_2018}
\begin{equation}
  p_{\sigma}(\phi)=
  \frac{1}{\sqrt{2\pi}\cdot\sigma} \exp\left(-\frac{\phi^2}{2\sigma^2}\right).
  \quad \phi\in \mathbb{R},
\end{equation}
The receiver is assumed to know the distribution $p_{\sigma}(\phi)$, but not the realization of $\phi$ for any given transmitted symbol.

Under a phase realization $\phi$, the two slot states in the S-PPM alphabet transform as
\begin{equation}
  \begin{aligned}
    &\text{symbol `0': } \ket{0,r}\to\ \ket{0,r;\phi} = R(\phi)\ket{0,r} = S(r\text{e}^{-2j\phi})\ket{0},\\
    &\text{symbol `1': } \ket{\alpha,r}\to\ \ket{\alpha,r;\phi} = R(\phi)\ket{\alpha,r} = D(\alpha \text{e}^{-j\phi})S(r\text{e}^{-2j\phi})\ket{0},
  \end{aligned}
\end{equation}
where $R(\phi) = \exp(-j\phi \hat{a}^\dagger \hat{a})$ is the single-mode phase-shift operator. 
Therefore, the $m$-th M-ary S-PPM codeword becomes
\begin{equation}
  \ket{\Psi_m(\phi)} = \bigotimes_{k=1}^{M} \ket{\psi_{m,k}(\phi)},
  \quad
  \ket{\psi_{m,k}(\phi)} = \begin{cases}
    D\!\left(\alpha e^{-i\phi}\right)S\!\left(r e^{-2i\phi}\right)\ket{0}, & k=m,\\[4pt]
    S\!\left(r e^{-2i\phi}\right)\ket{0}, & k\neq m.
  \end{cases}
\end{equation}
Since the receiver has no access to the actual phase realization, the relevant received state is the phase-averaged mixed state
\begin{equation}
  \rho_m^\text{pd} = \int_\mathbb{R} \text{d}\phi \cdot
  p_{\sigma}(\phi)\,
  \ket{\Psi_m(\phi)} \bra{\Psi_m(\phi)},
  \quad m=1,\dots,M.
  \label{eq:rho_pd_m}
\end{equation}

It is worth emphasizing that phase diffusion does not change the symbol energy, since $R(\phi)$ is a passive unitary operation. Hence the mean photon numbers of the empty slot and the pulse slot remain
$N_0=\sinh^2 r,\ N_1=\alpha^2+\sinh^2 r$
and the total mean photon number per $M$-ary S-PPM symbol is still $N=\alpha^2+M\sinh^2 r$.
Accordingly, the squeezing-energy fraction $\beta$ is unchanged by phase diffusion. 

In the following subsections, we first define the quantum benchmark, including the SRM benchmark used for high-order PPM, and the homodyne benchmark for the mixed-state ensemble $\{\rho_m^\text{pd}\}$, and then analyze the error probability of the IS-CPN receiver under the same phase-diffusion model.

\subsection{Quantum and homodyne benchmarks for S-PPM under phase diffusion}

Under the channel phase-diffusion model introduced above, the received $M$-ary S-PPM codewords are no longer pure states, but the mixed-state ensemble, as given by Eq.~\eqref{eq:rho_pd_m}. 
Therefore, the ideal pure-state Helstrom expression derived in Sec. II can no longer be directly applied. In this subsection, we define the quantum and the homodyne benchmarks for the phase-diffused S-PPM ensemble.

We first consider the quantum minimum-error benchmark. The Helstrom bound under phase diffusion is therefore given by \cite{Helstrom}
\begin{equation}
  P_\text{HB}^\text{S, pd}(\sigma)=1-
  \max_{\{\Pi_m\}}
  \frac{1}{M}
  \sum_{m=1}^{M}
  \Tr(\rho_m^\text{pd} \Pi_m),
  \label{eq:HB_phase_general}
\end{equation}
where the maximization is taken over all POVMs $\{\Pi_m\}_{m=1}^{M}$ satisfying
\begin{equation}
  \Pi_m \ge 0,
  \quad
  \sum_{m=1}^{M}\Pi_m = \mathbb{I}.
\end{equation}
For the binary case $M=2$, the exact Helstrom bound has the standard trace-norm form \cite{Olivares_phase_diffusion}
\begin{equation}
  P_\text{HB}^\text{S, pd}(\sigma)
  = \frac{1}{2}\left(1-\frac{1}{2}\left\|\rho_1^\text{pd}-\rho_2^\text{pd}\right\|_1\right),
\end{equation}
where \(\|A\|_1=\mathrm{Tr}\sqrt{A^\dagger A}\) denotes the trace norm. This formula can be evaluated numerically after truncating the single-mode Fock space.

For $M>2$, however, solving the exact mixed-state Helstrom optimization in the full multimode Fock space becomes computationally expensive.
The Hilbert-space dimension scales exponentially with $M$, and a direct semidefinite-program implementation rapidly becomes impractical. Therefore, for higher-order PPM, we use the square-root measurement associated with the phase-diffused ensemble as a tractable quantum benchmark. We denote the resulting error probability by $P_\text{SRM}^{\text{S,pd}}$.
The detailed construction of this SRM benchmark, including the use of the geometrically uniform symmetry and the efficient Gram-matrix method, is given in Appendix A. In the numerical results, the curves labeled ``S-PPM SRM'' refer to this SRM benchmark rather than to the exact Helstrom bound for $M>2$.

In contrast to the ideal phase-matched case, the phase-diffused quantum benchmark is no longer determined solely by the effective energy $N_\text{eff}$. It also depends explicitly on the phase-diffusion strength $\sigma$. As a result, the ideal analytical energy allocation $\beta_\text{opt}^\text{ideal}$ is no longer generally optimal. We therefore optimize the squeezing-energy fraction numerically under the fixed total-energy constraint. The fraction that minimizes the SRM benchmark is denoted by $\beta_\text{opt}^\text{pd}$. 
In the following subsection, we use the transmitted S-PPM signal designed with $\beta_\text{opt}^\text{pd}$. 

We next consider the homodyne receiver. 
For a given phase realization $\phi$, the empty-slot and pulse-slot hypotheses are
\begin{equation}
  H_0'(\phi):\ S\!\left(r e^{-2i\phi}\right)|0\rangle,
  \quad
  H_1'(\phi):\ D\!\left(\alpha e^{-i\phi}\right)S\!\left(r e^{-2i\phi}\right)|0\rangle.
\end{equation}
A homodyne measurement along the receiver x-quadrature gives two Gaussian conditional distributions,
\begin{equation}
  \begin{aligned}
    p(x|H_0',\phi) &= \frac{1}{\sqrt{2\pi V_\phi}}\exp\!\left(-\frac{x^2}{2V_\phi}\right),\\
    p(x|H_1',\phi) &= \frac{1}{\sqrt{2\pi V_\phi}}\exp\!\left[-\frac{(x-\alpha\cos\phi)^2}{2V_\phi}\right],
  \end{aligned}
\end{equation}
where the phase-dependent quadrature variance is $V_\phi = \frac{1}{4} \left( e^{-2r}\cos^2\phi + e^{2r}\sin^2\phi \right)$.
Since the receiver does not know the actual value of $\phi$, the effective homodyne distributions are obtained by averaging over the phase distribution:
\begin{equation}
  \begin{aligned}
    &\bar p(x|H_0') = \int_\mathbb{R} p_{\sigma_\phi}(\phi)\,p(x|H_0',\phi)\,\dd\phi,\\
    &\bar p(x|H_1') = \int_\mathbb{R} p_{\sigma_\phi}(\phi)\,p(x|H_1',\phi)\,\dd\phi.
  \end{aligned}
\end{equation}

Let $x_\text{th}$ be the single-slot decision threshold. The receiver declares the slot to be a pulse slot when $x>x_\text{th}$, and declares it to be an empty slot otherwise. The corresponding false-alarm and miss probabilities are
\begin{equation}
  \begin{aligned}
    &p_F(x_\text{th})=\int_{x_\text{th}}^{\infty}\bar p(x|H_0')\,\dd x=\int_\mathbb{R}p_{\sigma_\phi}(\phi)\,Q\!\left(\frac{x_\text{th}}{\sqrt{V_\phi}}\right)\dd\phi,\\
    &p_M(x_\text{th})=\int_{-\infty}^{x_\text{th}}\bar p(x|H_1')\,\dd x=\int_\mathbb{R}p_{\sigma_\phi}(\phi)\,Q\!\left(\frac{\alpha\cos\phi-x_\text{th}}{\sqrt{V_\phi}}\right)\dd\phi.
  \end{aligned}
\end{equation}
Unlike the ideal phase-matched case, these two probabilities are generally not equal. Therefore, the optimal homodyne threshold must be found numerically.

The $M$-ary decision is obtained using the same post-processing rule as in Sec. II. If exactly one slot is declared as the pulse slot, that slot is selected. If more than one slot is declared as the pulse slot, one of the declared slots is chosen uniformly at random. If no slot is declared as the pulse slot, one of all $M$ hypotheses is chosen uniformly at random.
For a fixed threshold $x_{th}$, the average correct-decision probability is
\begin{equation}
  \begin{aligned}
    P_{\text{cor},\sigma}^\text{Hom,S}(x_{th}) &= (1-p_M) \sum_{k=0}^{M-1} \binom{M-1}{k} p_F^{\,k}(1-p_F)^{M-1-k}\frac{1}{k+1} + \frac{1}{M}p_M(1-p_F)^{M-1}\\
    &= (1-p_M)\frac{1-(1-p_F)^M}{Mp_F} + \frac{1}{M}p_M(1-p_F)^{M-1}.
  \end{aligned}
\end{equation}
The homodyne benchmark under phase diffusion is then
\begin{equation}
  P_{\text{err},\sigma}^\text{Hom,S} = 1 - \max_{x_\text{th}} P_{\text{cor},\sigma}^\text{Hom,S}(x_{th}).
  \label{eq:SQL_phase_final}
\end{equation}

\subsection{IS-CPN error probability under phase diffusion}
We now analyze the symbol error probability of the proposed IS-CPN receiver in the presence of channel phase diffusion. 
As in Sec.~IV.A, we assume that all temporal slots within one M-ary PPM codeword experience the same random phase realization $\phi$, drawn from the probability density $p_\sigma(\phi)$. 
The receiver knows the distribution $p_\sigma(\phi)$, but not the actual phase realization for each received codeword.

In the ideal phase-matched case discussed in Sec. III, the inverse-squeezing operation maps the S-PPM alphabet exactly onto a coherent-state PPM alphabet.
Under phase diffusion, however, this exact mapping is no longer obtained.
After the channel phase rotation $R(\phi)$ and the receiver-side inverse-squeezing operation $S(-r)$, the empty-slot and pulse-slot states become
\begin{equation}
  \ket{\psi_0(\phi)} = U_{\phi}\ket{0},
  \quad
  \ket{\psi_1(\phi)} = D(\gamma_{\phi}) U_{\phi}\ket{0},
\end{equation}
where
\begin{equation}
  U_\phi = S(-r)R(\phi)S(r)
\end{equation}
is a residual Gaussian unitary caused by the mismatch between the phase-rotated squeezing ellipse and the fixed inverse-squeezing operation at the receiver.
The corresponding effective displacement is
\begin{equation}
  \gamma_\phi = \gamma\left(\cos\phi - i \text{e}^{-2r}\sin \phi\right), \gamma = \alpha \text{e}^r.
\end{equation}
Thus, under phase diffusion, the inverse-squeezing front end does not convert the received signal into an ordinary C-PPM signal. Instead, it produces a phase-dependent displaced residual Gaussian state.

For a given phase realization $\phi$, the thresholded no-click probability $q_T$ is rewritten from Eq.~\eqref{eq:no_click_prob_ideal} to
\begin{equation}
  Q_T\left(\kappa, \phi\right) = \Pr\left\{n<n_{th}|D(\kappa)U_\phi \ket{0}\right\} = \sum_{n=0}^{n_{th} - 1}\left|\bra{n}D(\kappa)U_\phi \ket{0}\right|^2.
\end{equation}
Therefore, the slotwise thresholded probabilities $p\left\{y_k|x_k^{\left(m\right)},u_k,\phi\right\}$ can be rewritten as
\begin{equation}
  \begin{array}{ll}
    p\left\{0|0,N,\phi\right\} = \Pr\left\{n<n_{th}|D(-\gamma)U_{\phi}\ket{0}\right\} = Q_T\left(-\gamma,\phi\right),
    & p\left\{1|0,N,\phi\right\} = 1 - Q_T\left(-\gamma,\phi\right),\\
    p\left\{0|1,N,\phi\right\} = \Pr\left\{n<n_{th}|D(-\gamma)D(\gamma_{\phi}) U_{\phi}\ket{0}\right\} = Q_T\left(\gamma_\phi-\gamma,\phi\right),
    & p\left\{1|1,N,\phi\right\} = 1-Q_T\left(\gamma_\phi-\gamma,\phi\right)\\
    p\left\{0|0,D,\phi\right\} = \Pr\left\{n<n_{th}|U_{\phi}\ket{0}\right\} = Q_T\left(0,\phi\right),
    & p\left\{1|0,D,\phi\right\} = 1-Q_T\left(0,\phi\right),\\
    p\left\{0|1,D,\phi\right\} = \Pr\left\{n<n_{th}|D(\gamma_{\phi}) U_{\phi}\ket{0}\right\} = Q_T\left(\gamma_\phi,\phi\right),
    & p\left\{1|1,D,\phi\right\} = 1 - Q_T\left(\gamma_\phi,\phi\right).
  \end{array}
\end{equation}
Conditioned on the common phase realization $\phi$, the likelihood of observing the complete thresholded record $\mathbf{y}$ under hypothesis $H_m$ is
\begin{equation}
  P\left(\mathbf{y}|H_m,\phi,n_{th}\right) = \prod_{k=1}^{M} 
  p\left\{y_k|x_k^{\left(m\right)},u_k\left(y_{<k}\right),\phi\right\}.
\end{equation}
Then the relevant likelihood is the phase-averaged likelihood
\begin{equation}
  \bar{P}\left(\mathbf{y}|H_m,n_{th}\right) = \int_{-\infty}^{\infty} \text{d}\phi \cdot p_\sigma\left(\phi\right) P\left(\mathbf{y}|H_m,\phi,n_{th}\right).
\end{equation}
For equiprobable hypotheses, the MAP rule is equivalent to the maximum-likelihood rule:
\begin{equation}
  \hat{m}\left(\mathbf{y}\right) \in \arg\max_{1\le m\le M} \bar{P}\left(\mathbf{y}|H_m,n_{th}\right).
\end{equation}
If several hypotheses attain the same maximum likelihood, one of the maximizers is chosen uniformly at random. With this tie-breaking convention, the average correct-decision probability is
\begin{equation}
  P_{\text{cor},\sigma}^{\text{IS-CPN}}\left(n_{th}\right) = \frac{1}{M} \sum_{\mathbf{y}\in\{0,1\}^M} \max_{1\le m \le M} \bar{P}\left(\mathbf{y}|H_m,n_{th}\right).
\end{equation}
Therefore, the symbol error probability of the thresholded-PNRD IS-CPN receiver under phase diffusion is
\begin{equation}
  P_{\text{err},\sigma}^{\text{IS-CPN}} = 1 - \max_{n_{th}\ge 1} P_{\text{cor},\sigma}^\text{IS-CPN}\left(n_{th}\right).
\end{equation}
This expression provides a finite-path evaluation of the IS-CPN receiver under the common-phase-diffusion model. Since each temporal slot produces a binary thresholded outcome, the number of complete detection records is $2^M$.

As a consistency check, we consider the phase-matched limit $\phi\to 0, \sigma\to 0$. In this limit, $U_\phi \to \mathbb{I}$, $\gamma_\phi\to \gamma$. Consequently, $Q_T(\gamma, 0) = q_T$ and $Q_T(0, 0) = 1$. Substituting these probabilities into the finite-path MAP formulation recovers the ideal error probability (Eq.~\eqref{eq:is-cpn_ideal}). Hence, the present formulation is a direct generalization of the ideal IS-CPN analysis to the phase-diffusion case.

\section{Numerical results and discussion}
In this section, we numerically evaluate the performance of the proposed inverse-squeezing conditional pulse-nulling receiver and compare it with conventional coherent-state PPM receivers and the benchmark limits derived in the previous sections. The photon number per bit is defined as
\begin{equation}
  N_b = \frac{N}{\log_2 M}\ \text{bit}^{-1}
\end{equation}
where $N$ is the total mean photon number per $M$-ary PPM symbol. Unless otherwise specified, all error probabilities shown below are symbol error probabilities.
In the numerical simulations, single-mode Gaussian states and photon-number distributions were evaluated in a truncated Fock basis using QuTiP \cite{qutip}, while the SRM benchmarks were computed using the block Gram-matrix method described in Appendix A. The photon-number cutoff was chosen sufficiently large to ensure convergence; increasing the cutoff produced negligible changes in the plotted error probabilities.

\begin{figure}[htbp]
  \centering
  \includegraphics[scale=1.0]{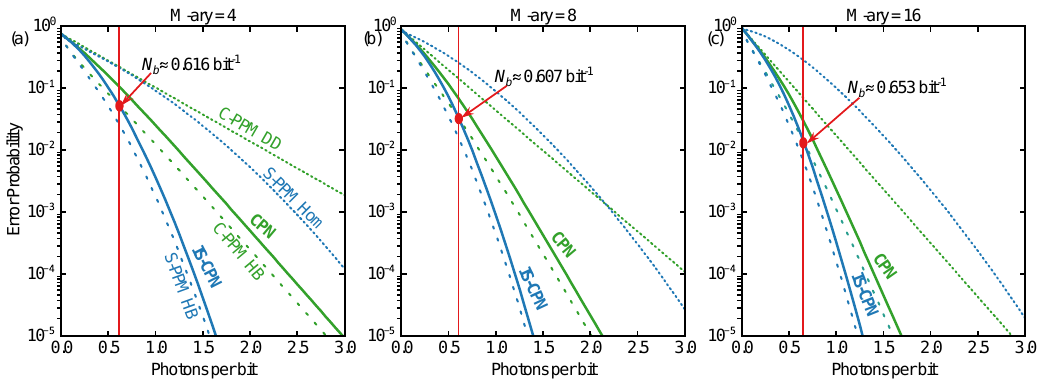}
  \caption{In the ideal phase-matched case, the performance of the inverse-squeezing conditional pulse nulling receiver (IS-CPN) as a function of the photons per bit.}
  \label{fig:ideal_error_prob}
\end{figure}

We first consider the ideal case. Figure~\ref{fig:ideal_error_prob} shows the error probabilities of the proposed IS-CPN receiver and the conventional CPN receiver for M-ary PPM as functions of photons per bit. The corresponding C-PPM and S-PPM benchmarks are also shown for comparison.
In the absence of phase diffusion, the inverse-squeezing front end exactly maps the S-PPM alphabet onto a coherent-state PPM alphabet with the enhanced pulse energy $N_\text{eff} = N\left(1 + \frac{N}{M}\right)$. Therefore, the IS-CPN receiver has the same functional form as the conventional CPN receiver, but with the original coherent pulse energy replaced by $N_\text{eff}$. This effective-energy enhancement is the main reason for the performance improvement observed in Fig.~\ref{fig:ideal_error_prob}.

The IS-CPN receiver consistently outperforms the conventional CPN receiver under the same total-energy constraint. Moreover, in the plotted energy range, the IS-CPN curve can even fall below the C-PPM Helstrom bound evaluated at the same total energy. This does not contradict the Helstrom limit, because the C-PPM Helstrom bound is the quantum limit for discriminating the coherent-state PPM ensemble, whereas the proposed receiver operates on a different transmitted ensemble, namely S-PPM. The improvement is therefore a consequence of signal-state engineering: part of the total energy is allocated to squeezing, which increases the effective displacement after inverse squeezing.
For the examples shown in Fig.~\ref{fig:ideal_error_prob}, the IS-CPN receiver crosses the C-PPM Helstrom benchmark at approximately $N_b=0.616,\ 0.607,\ \text{and}\ 0.653\ \text{bit}^{-1}$ for the corresponding PPM orders shown in panels (a)-(c), respectively. These crossing points provide a direct illustration of the benefit of the inverse-squeezing-enhanced displacement in ideal condition.

\begin{figure}[htbp]
  \centering
  \includegraphics[scale=1.0]{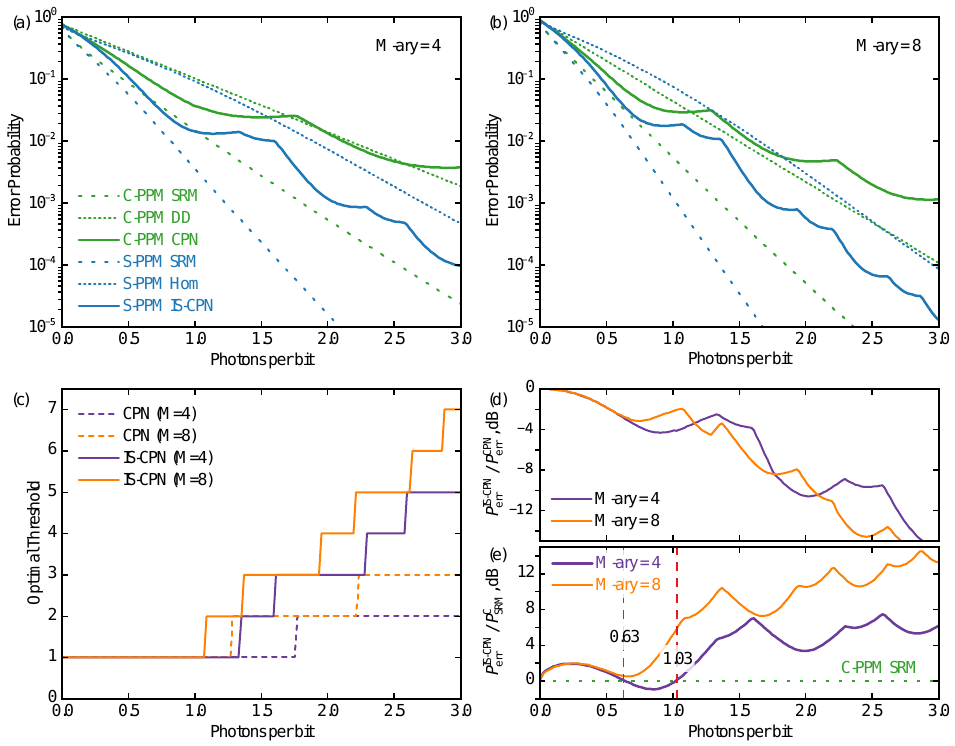}
  \caption{Under phase-diffusion condition ($\sigma = 0.1$), the performance of the inverse-squeezing conditional pulse nulling receiver (IS-CPN) versus the photons per bit. (a) and (b) error probabilities of $M=4,\ 8$; (c) the optimal threshold $n_{th}^*$; (d) and (e) the ratios $P_\text{err}^\text{IS-CPN} / P_\text{err}^\text{CPN}$ and $P_\text{err}^\text{IS-CPN} / P_\text{SRM}^\text{C}$ (dB).}
  \label{fig:pd_error_prob}
\end{figure}
We next evaluate the receiver under the phase-diffusion model introduced in Sec.~IV.A.
Figure~\ref{fig:pd_error_prob} shows the performance for $\sigma = 0.1$. Panels (a) and (b) compare the error probabilities of IS-CPN, conventional CPN, the C-PPM DD benchmark, the C-PPM SRM benchmark, the S-PPM SRM benchmark, and the S-PPM homodyne benchmark for $M=4$ and $M=8$.
Compared with the ideal case, phase diffusion degrades the benefit of inverse squeezing because the received squeezing ellipse is no longer perfectly matched to the fixed inverse-squeezing operation at the receiver. Nevertheless, Fig.~\ref{fig:pd_error_prob} shows that a significant performance advantage remains. The IS-CPN receiver lies below the conventional CPN receiver over the plotted energy range. Moreover, the IS-CPN can fall below the C-PPM SRM curve in the interval $N_b \in (0.63,1.03)$ (see panel (e)). This indicates that the inverse-squeezing operation still provides a useful effective-displacement enhancement under moderate phase diffusion.

The curves also show small step-like and oscillatory features. These features arise from the discrete switching of the optimized photon-number threshold $n_{th}$, as shown in Fig.~\ref{fig:pd_error_prob}(c).
Fig.~\ref{fig:pd_error_prob}(d) shows the ratio $P_\text{err}^\text{IS-CPN}/P_\text{err}^\text{CPN}$ (dB). Negative values indicate that IS-CPN outperforms conventional CPN. The ratio remains below 0 dB throughout the plotted range, and becomes more pronounced at higher photon numbers. In the range $N_b \approx 2.0$ to 3.0, the error probability of IS-CPN is roughly one order of magnitude or more below that of CPN.

\begin{figure}[htbp]
  \centering
  \includegraphics[scale=1.0]{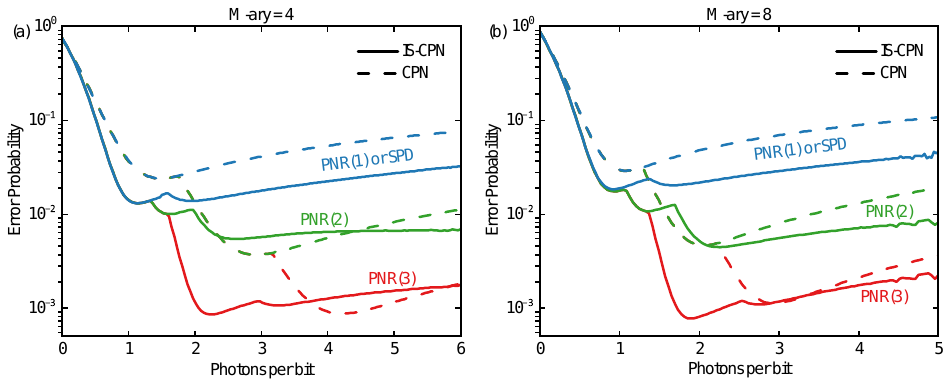}
  \caption{Under phase-diffusion condition ($\sigma = 0.1$), the performance of IS-CPN and CPN as a function of the photons per bit for different photon-number resolution. PNR(K) denotes the case in which the threshold is optimized over $n_{th} = 1, \cdots, K$.}
  \label{fig:pd_error_prob_resolution}
\end{figure}
Figure~\ref{fig:pd_error_prob_resolution} investigates the effect of finite photon-number resolution under phase diffusion with $\sigma = 0.1$. Here PNR(K) denotes the case in which the threshold is optimized over $n_{th} \in \left\{1,\cdots,K\right\}$. Thus, PNR(1) corresponds to a single-photon detector (SPD), while larger $K$ allows the receiver to exploit higher photon-number thresholds.

Both IS-CPN and CPN exhibit a characteristic saturation or rebound at high photon numbers when the available photon-number resolution is limited. That is because, as the signal energy increases, the optimal threshold generally shifts to larger photon numbers. If the detector resolution or the allowed threshold range is insufficient, the receiver can no longer track the unconstrained optimum. As a result, the error probability stops decreasing and may even increase.
The rebound is more pronounced for conventional CPN than for IS-CPN. This indicates that the inverse-squeezing receiver is less sensitive to the threshold limitation in the considered regime. For PNR(1), IS-CPN consistently outperforms CPN for both $M=4$ and $M=8$. For higher photon-number resolutions, CPN can be slightly better in certain intermediate photon-number intervals. However, because the CPN curve rebounds more strongly at higher energies, IS-CPN eventually regains the advantage.

\begin{figure}[htbp]
  \centering
  \includegraphics[scale=1.0]{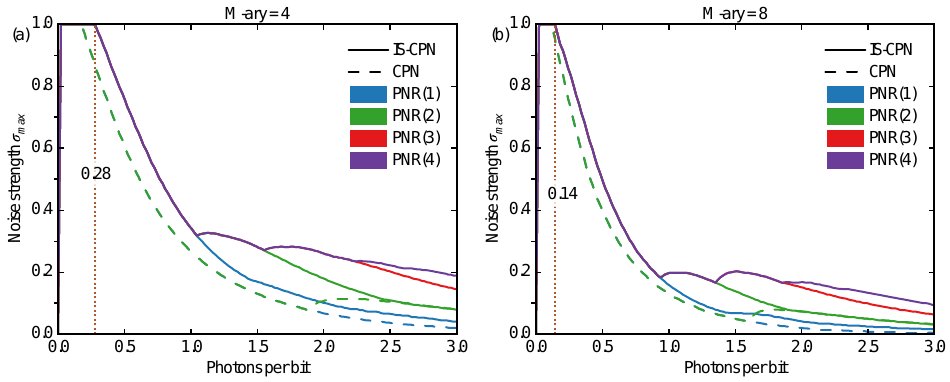}
  \caption{Maximum tolerable phase noise $\sigma_{max}$ of IS-CPN and CPN as a function of the photons per bit for different photon number resolution. The region below each curve corresponds to the parameter regime where the receiver outperforms C-PPM DD.}
  \label{fig:pd_max_sigma}
\end{figure}
To quantify robustness against phase diffusion more directly, we define the maximum tolerable phase-diffusion strength $\sigma_\text{max}$. For a given receiver and photon number per bit $N_b$, $\sigma_\text{max}$ is defined as
\begin{equation}
  \sigma_\text{max}(N_b) = \sup \left\{\sigma:\ P_\text{err}^\text{receiver}(N_b, \sigma)<P_\text{err}^\text{C, DD}(N_b, \sigma)\right\}
\end{equation}
In other words, $\sigma_\text{max}$ is the largest phase-noise strength for which the receiver still outperforms the direct-detection benchmark of ordinary coherent-state PPM under the same total-energy constraint.

Figure~\ref{fig:pd_max_sigma} plots $\sigma_\text{max}$ for IS-CPN and conventional CPN as functions of photons per bit for different photon-number resolutions. The region below each curve corresponds to the parameter regime in which the corresponding receiver beats the C-PPM DD benchmark. The IS-CPN receiver exhibits a larger $\sigma_\text{max}$ than conventional CPN over the plotted range. This means that the inverse-squeezing receiver can tolerate stronger phase diffusion while still maintaining an advantage over C-PPM direct detection. The effect of photon-number resolution is also visible in Fig.~\ref{fig:pd_max_sigma}. Increasing the allowed threshold range generally increases $\sigma_\text{max}$, but the improvement is not uniform over all photon numbers. At low photon numbers, on-off detection is already close to optimal, so higher resolution provides only a limited benefit. At moderate and high photon numbers, however, larger thresholds become important, and the advantage of higher photon-number resolution becomes more visible.

\begin{figure}[htbp]
  \centering
  \includegraphics[scale=1.0]{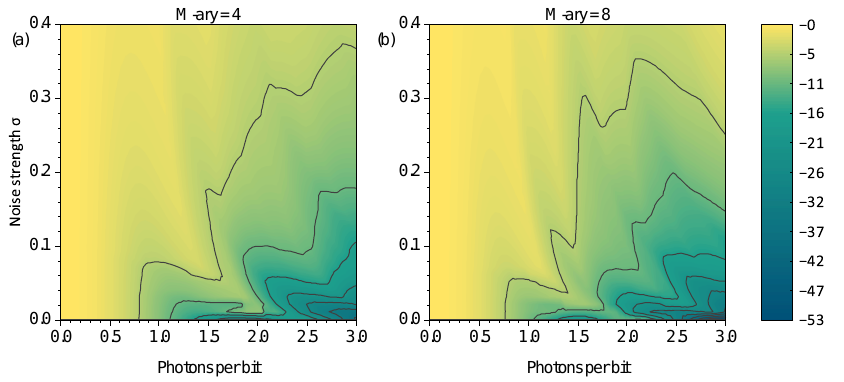}
  \caption{Under phase-diffusion condition, the ratio $P_\text{err}^\text{IS-CPN} / P_\text{err}^\text{CPN}$ (dB) as a function of the noise strength $\sigma$ and the photons per bit, with $n_\text{th}$ optimized over $1,\cdots,5$.}
  \label{fig:pd_contour}
\end{figure}
Finally, Fig.~\ref{fig:pd_contour} provides a two-dimensional view of the performance gain of IS-CPN over conventional CPN. The plotted quantity is the ratio $P_\text{err}^\text{IS-CPN}\left(N_b,\sigma\right) / P_\text{err}^\text{CPN}\left(N_b,\sigma\right)$ (dB). A negative value of the ratio indicates that IS-CPN outperforms CPN, while a positive value indicates that CPN performs better. The contour plots show that IS-CPN provides a broad performance gain over CPN in the considered $(N_b,\sigma)$ parameter region. A significant portion of the plotted region exhibits a gain exceeding 5 dB.

Overall, the numerical results show that inverse squeezing provides two related benefits. In the ideal case, it converts S-PPM into an equivalent C-PPM signal with enhanced effective energy $N_\text{eff}$. Under phase diffusion, this exact equivalence is lost, but the receiver still preserves part of the effective-displacement advantage. The finite-threshold and robustness analyses further show that IS-CPN is not only theoretically advantageous, but also more robust than conventional CPN over a broad practical parameter regime.

\section{Conclusion}
In this paper, we investigated squeezed-state pulse-position modulation and inverse-squeezing receivers under both ideal and phase-diffusion conditions. In the proposed S-PPM format, the empty slots are squeezed vacuum states and the pulse slot is a displaced squeezed state. Under ideal phase matching, the inverse-squeezing front end exactly converts S-PPM into an equivalent coherent-state PPM signal with enhanced pulse energy $N_\text{eff}$. This allows the proposed IS-CPN receiver to inherit the structure of conventional CPN while operating on a larger effective displacement, leading to a closed-form expression for the ideal error probability.
 
We further analyzed the receiver under a common phase-diffusion model. In this case, inverse squeezing no longer produces an ordinary coherent-state PPM signal, because a residual Gaussian unitary remains after the inverse-squeezing operation. To evaluate the receiver in this regime, we developed a finite-path MAP formulation based on thresholded photon-number records and phase-averaged likelihoods. Numerical results show that IS-CPN retains a significant advantage over conventional CPN under moderate phase diffusion. The receiver also exhibits improved tolerance to phase noise and finite photon-number resolution while maintaining an advantage over the C-PPM direct-detection benchmark.

The present analysis uses a thresholded-record MAP rule, in which the photon-number-resolving detector output is first converted into a binary click/no-click sequence according to the threshold $n_{th}$. This choice is consistent with the switching logic of the conditional pulse-nulling receiver and gives an exact finite-path expression with only $2^M$ records. A natural extension is to retain the complete photon-number record for the final MAP decision while keeping the same sequential nulling/direct-detection policy. Since the thresholded record is a coarse-grained version of the full photon-number outcome, such a full-record MAP post-processing strategy is expected to perform no worse, and may further improve the robustness of IS-CPN under phase diffusion and other nonideal conditions. Its implementation, however, requires summation over a much larger photon-number outcome space and is therefore left for future work.

\appendix

\section{Quantum benchmark evaluation for phase-diffused S-PPM: GUS reduction and SRM approximation}
In the presence of channel phase diffusion, the received $M$-ary S-PPM codewords form the mixed-state ensemble $\left\{\rho_m^\text{pd}\right\}^M_{m=1}$ defined in Eq.~\eqref{eq:rho_pd_m}.
For $M=2$, the exact Helstrom bound can be evaluated through the trace-norm formula given in Sec.~IV.B. For higher-order PPM, however, directly solving the exact mixed-state Helstrom optimization in the full multimode Fock space becomes computationally expensive.

This appendix explains how the quantum benchmark used in Sec. IV.B is evaluated.
We first show that the common phase-diffusion model preserves the geometrically uniform symmetry of the PPM ensemble, which reduces the number of independent POVM operators in the exact Helstrom problem.
Since this reduction does not remove the exponential scaling of the multimode Hilbert space, we then introduce the square-root measurement (SRM) benchmark used for the high-order numerical results. Finally, we present an efficient Gram-matrix construction that avoids explicitly building the full multimode tensor-product states.

\subsection{GUS reduction of the phase-diffused ensemble}
The geometrically uniform symmetry (GUS) structure of the PPM, whether C-PPM or S-PPM, constellation is apparent \cite{ppm_hb_numerical_problems}. Formally, the state $\rho_{i+1}$ is obtained from $\rho_i$ through a symmetry operator $T$ that pushes each Kronecker factor by one position
(modulo M) to the left.
\begin{equation}
  \rho_m \sim \ket{\Psi_m} = T^{m-1} \ket{\Psi_1},\ m=1,\cdots,M.
\end{equation}
Under the common--phase--diffusion model adopted in Sec. IV.A, all temporal slots within one symbol experience the same random phase realization. Hence, for a given phase $\phi$, the phase rotation acting on the whole codeword can be written as
\begin{equation}
  \mathcal{R}_\phi = R(\phi)^{\otimes M},
\end{equation}
and the phase-rotated codeword is
\begin{equation}
  \ket{\Psi_m(\phi)} = \mathcal{R}_\phi\ket{\Psi_m}.
\end{equation}

Since $T$ merely rearranges the slots, its action on the mode operators is
\begin{equation}
  T a_k T^\dagger = a_{k+1}.
\end{equation}
Hence
\begin{equation}
  T\left(\sum_{k=1}^{M}a_k^\dagger a_k\right) T^\dagger = \sum_{k=1}^{M} T a_k^\dagger a_k T^\dagger = \sum_{k=1}^{M}a_{k+1}^\dagger a_{k+1} = \sum_{k=1}^{M}a_k^\dagger a_k.
\end{equation}
Then
\begin{equation}
  T\mathcal{R}_\phi T^\dagger = T \exp\left(-i\phi \sum_{k=1}^{M} a^\dagger_k a_k\right) T^\dagger = \exp\left(-i\phi\cdot T \left(\sum_{k=1}^{M} a^\dagger_k a_k\right) T^\dagger \right) = \mathcal{R}_\phi.
\end{equation}
Consequently
\begin{equation}
  \left[T, \mathcal{R}_\phi\right] = 0.
\end{equation}
Therefore,
\begin{equation}
  \ket{\Psi_m(\phi)} = T^{m-1}\ket{\Psi_1(\phi)},
\end{equation}
and after averaging over the unknown phase we obtain
\begin{equation}
  \rho_m^\text{pd} = T^{m-1} \rho_1^\text{pd} T^{-(m-1)}.
\end{equation}
Thus, even under phase diffusion, the received S-PPM ensemble remains a symmetric mixed-state PPM constellation of GUS type. For equiprobable hypotheses, the exact Helstrom benchmark involves M POVM operators $\left\{\Pi_m\right\}_{m=1}^M$. By the standard GUS argument, the optimal POVM can be restricted to the covariant form \cite{ppm_hb_numerical_problems}
\begin{equation}
  \Pi_m = T^{m-1} \Pi_1 T^{-(m-1)},\ m=1,\cdots,M.
\end{equation}
Accordingly, the M-operator optimization can be reduced to a single reference POVM element:
\begin{equation}
  P_{\text{HB}}^\text{S,pd} = 1 - \max_{\Pi_1} \Tr(\rho_1 \Pi_1),
\end{equation}
subject to $\Pi_1 \ge 0$ and $\sum_{m=1}^{M} T^{m-1} \Pi_1 T^{-(m-1)} = \mathbb{I}$. 
This covariance reduction decreases the number of independent POVM variables. However, it does not remove the exponential growth of the multimode Hilbert-space dimension, which scales as $(N_\text{cut})^{M}$ after Fock-space truncation. Therefore, for the high-order numerical results, we use the SRM benchmark described below.

\subsection{SRM benchmark for high-order phase-diffused PPM}
To obtain a tractable quantum benchmark for high-order phase-diffused S-PPM, we employ the square-root measurement associated with the mixed-state ensemble $\{\rho_{m}^{pd}\}_{m=1}^{M}$. This measurement is not assumed to be the exact Helstrom measurement for general mixed-state ensembles; rather, it provides a physically meaningful and computationally efficient quantum benchmark.

For each hypothesis, let 
\begin{equation}
  \rho_{m}^{pd}=\zeta_{m}\zeta_{m}^{\dagger}
\end{equation}
be a factorization of the density operator, and define the state matrix 
\begin{equation}
  Z=[\zeta_{1},\zeta_{2},\cdot\cdot\cdot,\zeta_{M}].
\end{equation}
The corresponding block Gram matrix is defined as:
\begin{equation}
  G=Z^{\dagger}Z, 
\end{equation}
whose $(i,j)$-th block is $G_{ij}=\zeta_{i}^{\dagger}\zeta_{j}$.
The SRM measurement operator associated with hypothesis $H_m$ is formally constructed as 
\begin{equation}
  \Pi_{m}^\text{SRM}=\mu_{m}\mu_{m}^{\dagger}, 
\end{equation}
where $\mu_{m} = Z(G^{-1/2})_{\bullet m}$ is the $m$-th block column of $ZG^{-\frac{1}{2}}$. Therefore, the error probability is given by
\begin{equation}
  P_{\text{err,SRM}}^\text{S,pd} = 1 - \frac{1}{M}\sum_{m=1}^{M}\Tr\left(\rho_m^\text{pd}\Pi_m^\text{SRM}\right)
\end{equation}
This expression can be evaluated directly from the Gram matrix. Indeed,
\begin{equation}
  \Tr\left(\rho_{m}^\text{pd}\Pi_{m}^\text{SRM}\right) = \Tr(\zeta_{m}\zeta_{m}^{\dagger}\mu_{m}\mu_{m}^{\dagger}) = \Tr(\zeta_{m}^{\dagger}\mu_{m} \mu_{m}^{\dagger}\zeta_{m}) = \left\|\zeta_m^\dagger \mu_m\right\|_F^2,
\end{equation}
where $||\cdot||_F$ denotes the Frobenius norm. Using the definition of $\mu_m$, we have 
\begin{equation}
  \zeta_m^\dagger \mu_m = \zeta_m^\dagger Z (G^{-1/2})_{\bullet m}.
\end{equation}
The factor $\zeta_m^\dagger Z$ is the $m$-th block row of $G$. Hence
\begin{equation}
  \zeta_m^\dagger \mu_m = G_{m \bullet} (G^{-1/2})_{\bullet m} = (G^{1/2})_{m m}
\end{equation}
Therefore,
\begin{equation}
  \Tr\left(\rho_{m}^\text{pd}\Pi_{m}^\text{SRM}\right) = \left\|\left(G^{\frac{1}{2}}\right)_{m m} \right\|_F^2.
\end{equation}
The SRM error probability becomes
\begin{equation}
  P_{\text{err,SRM}}^\text{S,pd} 
  = 1 - \frac{1}{M}\sum_{m=1}^{M}\left\|\left(G^{\frac{1}{2}}\right)_{m m} \right\|_F^2
\end{equation}

\subsection{Efficient Numerical Construction of the Gram Matrix}

Although the SRM error probability has been analytically reduced to operations on the Gram matrix $G$, constructing $G$ explicitly still suffers from the ``curse of dimensionality" because the inner products $\zeta_{i}^{\dagger}\zeta_{j}$ reside in an exponentially large tensor-product space $(N_{cut})^M$. Here we introduce an efficient numerical method to bypass this multi-mode tensor space entirely.

Under the common--phase--diffusion model, we approximate the continuous integral of the mixed state (Eq.~\eqref{eq:rho_pd_m}) using the Gauss-Hermite quadrature, discretizing the phase into $K$ nodes $\phi_{k}$ with corresponding weights $w_{k}$:
\begin{equation}
  \rho_m^\text{pd} \approx \sum_{k=1}^K w_k |\Psi_m(\phi_k)\rangle\langle\Psi_m(\phi_k)|
\end{equation}
With this discretization, the state matrix $\zeta_{m}$ is explicitly composed of $K$ weighted pure-state column vectors:
\begin{equation}
  \zeta_m = \big[ \sqrt{w_1}|\Psi_m(\phi_1)\rangle, \sqrt{w_2}|\Psi_m(\phi_2)\rangle, \cdots, \sqrt{w_K}|\Psi_m(\phi_K)\rangle \big]
\end{equation}
The matrix $G$ is therefore of size $MK\times MK$, and its $K\times K$ sub-block $G_{mn}$ corresponds to the overlap between the m-th and n-th hypotheses. Exploiting the geometrically uniform tensor-product structure of the PPM codewords, the multi-mode inner product strictly factors into a product of single-mode inner products:
\begin{equation}
  \langle\Psi_m(\phi_i)|\Psi_n(\phi_j)\rangle = \prod_{s=1}^M \langle\psi_{m,s}(\phi_i)|\psi_{n,s}(\phi_j)\rangle
\end{equation}
Consequently, the $(i, j)$-th element of the sub-block $G_{mn}$ is computed as:
\begin{equation}
  [G_{mn}]_{ij} = \sqrt{w_i w_j} \prod_{s=1}^M \langle\psi_{m,s}(\phi_i)|\psi_{n,s}(\phi_j)\rangle
\end{equation}
By doing so, the full block Gram matrix $G$ is assembled purely through element-wise scalar exponentiation and multiplication of pre-computed single-mode inner products. Combined with the algebraic reduction derived in Appendix~A.2, this method achieves near-instantaneous execution times and a minimal memory footprint for evaluating the phase-diffused quantum limit.

\begin{acknowledgments}
This work was supported by 
Guangxi Science and Technology Program (Grant No.~GuiKeFN2600640534),
Innovative Talent Development Fund of Information Support Force Engineering University (Grant No.~XJKT-QT-25-02-GW, No.~XJKT-QT-25-03-GW),
and Guangxi Science and Technology Base and Talent Project (Grant No.~GuiKeAD25069071).
\end{acknowledgments}

\section*{Data Availability}
The data that support the findings of this article are not publicly available. The data are available from the authors upon reasonable request.

\nocite{*}

\bibliography{my_citation}% Produces the bibliography via BibTeX.

\end{document}